\documentclass[10pt]{iopart}
\usepackage{mathrsfs} 
\usepackage[usenames,dvipsnames,svgnames,table]{xcolor}

\begin{document}

\title{The Algebraic-Hyperbolic Approach to the Linearized Gravitational Constraints on a Minkowski
Background}
\author{Jeffrey Winicour$^{1,2}$}

\address{
${}^{1}$ Department of Physics and Astronomy \\
        University of Pittsburgh, Pittsburgh, PA 15260, USA\\
${}^{2}$ Max-Planck-Institut f\" ur
         Gravitationsphysik, Albert-Einstein-Institut, \\
	 14476 Golm, Germany \\
	}

\begin{abstract}

An algebraic-hyperbolic method for solving the Hamiltonian and momentum constraints
has recently been shown to be well posed for general nonlinear perturbations of the
initial data for a Schwarzschild black hole.
This is a new approach to solving the constraints of Einstein's equations which does not involve elliptic equations and has potential importance for the construction of binary
black hole data.
In order to shed light on the underpinnings of this approach, we consider its application
to obtain solutions of the constraints for linearized
perturbations of Minkowski space. In that case, we find the surprising result
that there are no suitable Cauchy hypersurfaces in Minkowski space
for which the linearized algebraic-hyperbolic constraint problem  is well posed.

\end{abstract}


\maketitle

\bigskip

\section{Introduction}

Physically realistic initial Cauchy data for Einstein's equations are a
major ingredient for the
numerical simulation of gravitational systems such as binary black holes.
The formulation of such initial data is further complicated mathematically
by the nonlinear constraint equations that they must satisfy. The traditional approach
expresses the constraints in the form of elliptic equations,
based upon the conformal treatment of the Hamiltonian constraint
introduced by Lichnerowicz~\cite{lich}
and later extended by York~\cite{york0,york1} to treat the momentum
constraint. For reviews see~\cite{cook,gourg}.
Recently, a new method of
solving the constraints which only involves  algebraic and hyperbolic equations~\cite{racz1}
has been shown to lead to
a well-posed constraint problem in the case of
nonlinear perturbations of Schwarzschild black hole data~\cite{raczwin1}.
A well-posed problem is a necessity for a stable numerical simulation.

The details of the gravitational
waveform supplied by numerical simulation of the inspiral and merger of  a binary black hole
is a key input for interpreting the scientific content of the observed
signal. It is thus important that the initial data does
not introduce spurious effects into the waveform, in particular the ``junk radiation''
common to current elliptic formulations of the constraint problem.
Elliptic equations require boundary data at inner boundaries
in the strong field region
surrounding the singularities inside the black holes, as well as at the outer boundary
in the far field of the system. The treatment of the inner boundary
is a potential source of junk radiation. The algebraic-hyperbolic method only requires
data on the outer boundary, where the choice of
boundary data can be guided by asymptotic flatness.
The constraints are then satisfied by an  inward ``evolution''
of the hyperbolic system along radial streamlines of the initial Cauchy hypersurface.
The possibility of extending this method to binary black holes offers an alternative
approach to the initialization problem that might prove to be more physically realistic.

In this note, in order to shed further light on the algebraic-hyperbolic
formulation of the constraint system 
we investigate its application to linearized perturbations of
initial data on a Minkowski background. We are led to the surprising result that there are no {\em useful}
Cauchy hypersurfaces in Minkowski spacetime for which this linearized constraint problem
is well posed.

By {\em useful}, we require that the Cauchy data be 
asymptoically flat so it can be expressed in non-singular
form on a Cauchy hypersurface foliated by topological 2-spheres. The center of such a foliation has
a singular limit where the area of the 2-sphere is zero. In the case of
data for a perturbed Schwarzschild black hole, the singular point of the foliation agrees with the
spacetime singularity and is therefore removed from the Cauchy hypersurface.
In the perturbed Minkowski case, we require smoothness of the data at the center when
referred to a Cartesian coordinate system.
This leads to the requirements for a {\em useful} Cauchy hypersurface and foliation
given in Sec.~\ref{sec:lin}. Our results indicate
that the algebraic-hyperbolic approach is applicable
to asymptotically flat data only for spacetimes with singularities.

The initial data for solving Einstein's equations consist of a pair of symmetric tensor fields
$(h_{ab},k_{ab})$
on a smooth space-like three-dimensional manifold $\Sigma$, where $h_{ab}$ is a Riemannian metric and
$k_{ab}$ is interpreted as the extrinsic curvature of $\Sigma$  after its embedding
in a 4-dimensional space-time.
The constraints on initial data for a vacuum solution (see e.g.~\cite{choquet,wald})
consist of the Hamiltonian constraint
\begin{equation}
     {}^{(3)} R  +({k^a}_a)^2 - k_{ab} k^{ab} =0\,,
      \label{eq:ham}
\end{equation}     
and the momentum constraints
\begin{equation}
    D_b {k^{b}}_{a} - D_a k =0\, , \quad k= {k^{b}}_{b} \,,
    \label{eq:mom}
\end{equation}
where ${}^{(3)}R$ and $D_a$ denote the scalar curvature and the
covariant derivative associated with $h_{ab}$. 

The approach to solving the constraints depends upon the
choice of which metric and extrinsic curvature
variables are prescribed freely. In the algebraic-hyperbolic approach, the free variables
consist of the initial 3-metric $h_{ab}$ and four components of the extrinsic
curvature which are picked out by a foliation of $\Sigma$ by topological 2-spheres.
In the case where $\Sigma$ is spherical symmetric with foliation by metric
2-spheres of surface area $4\pi r^2$, so that 
\begin{equation}
 h_{ab}=R_a R_b + Q_{ab}, \quad Q_{ab} =r^2 q_{ab}, \quad R_a= D_a r( h^{bc} D_b r D_c r)^{-1/2},
 \label{eq:sphdec}
\end{equation}
where $q_{ab}$ is the unit sphere metric,
the freely prescribed components
of the extrinsic curvature are the transverse-tracefree components
\begin{equation}
   K^{(TT)}_{ab} =Q_a^c Q_b^d k_{cd}-\frac {1}{2} Q_{ab} Q^{cd} k_{cd}.
\label{eq:TTK}
\end{equation}
The remaining 4 components of the extrinsic curvature are then
determined by the constraints. The component 
\begin{equation}
\kappa =  R^c R^d k_{cd}
\label{eq:kappa}
\end{equation}
is determined algebraically from the Hamiltonian constraint and
the components
\begin{equation}
K_a= R^b k_{ab} - \kappa R_a \quad \mbox{and} \quad K= Q^{ab} k_{ab}
\label{eq:Ka}
\end{equation}
are then determined by the momentum constraints.
In this process, the momentum constraints form a well-posed
symmetric hyperbolic system for $(K,K_a)$ provided that the inequality
\begin{equation}
               \kappa K < 0
               \label{eq:ineq}
\end{equation}
is satisfied.
For general nonlinear perturbations of a Schwarzschild spacetime,
it has been shown that this inequality is satisfied~\cite{raczwin1} and a pseudo-code
for numerical implementation of the algebraic-hyperbolic constraint
system has been formulated~\cite{raczwin2}.

This algebraic-hyperbolic approach was introduced by R\'acz
along with other possibilities for solving
the constraints without elliptic equations~\cite{racz1,racz2}. One of those possibilities
is a parabolic-hyperbolic formulation, in which the Hamiltonian
constraint is reduced to a parabolic equation. In subsequent work,
R\'acz has shown that the parabolic-hyperbolic system has attractive
features for constructing binary black hole data~\cite{racz3}. It has also been show that
the parabolic-hyperbolic constraint problem is well posed for
nonlinear perturbations of Minkowski space, and current work has successfully
carried out a numerical implementation of
this problem~\cite{schell}. Here we investigate the application of the algebraic-hyperbolic
approach to perturbed Minkowski data. 

\section{Linearized algebraic-hyperbolic constraint system on a Minkowski background}
\label{sec:lin}

Consider the linearized perturbation 
$ (\delta h_{ab},\delta k_{ab})$ of data for Minkowski space, with background metric
$\eta_{ab} =diag(-,+,+,+)$
in standard Cartesian inertial coordinates $x^a= (t,x^i) =(t,x,y,z)$.
For the perturbed algebraic-hyperbolic constraint system to be well posed
the inequality (\ref{eq:ineq}) must be satisfied. However, for a Cauchy hypersurface $\Sigma$
based on the standard inertial time slicing, the extrinsic curvature $k_{ab}$ of the
$t=constant$ hypersurfaces vanishes so that (\ref{eq:ineq}) cannot be satisfied
to linearized order. Moreover, the Hamiltonian constraint (\ref{eq:ham}) can be re-expressed
in the form
\begin{equation}
      2\kappa K = 2 K^a K_a -\frac{1}{2} K^2 - {}^{(3)} R 
      +K^{(TT)ab} K^{(TT)}_{ab},
 \end{equation} 
where the components $(\kappa,K,K_a,K^{(TT)}_{ab})$ of the extrinsic curvature $k_{ab}$ are defined
with respect to the foliation of $\Sigma$ in a manner analogous to (\ref{eq:TTK}) -- (\ref{eq:Ka}) for the
case of a spherically symmetric metric.
For a generic perturbation of the Minkowski background
$\delta  {}^{(3)} R \ne 0$ so that
$K\delta \kappa+\kappa\delta K \ne 0$ to linear order. where $K$ and $\kappa$
refer to the background. Thus
the stability inequality (\ref{eq:ineq}) 
cannot be satisfied to linear order unless the Cauchy hypersurface for the
Minkowski background be chosen to have nonzero extrinsic curvature. 

This leads to the question whether there are Cauchy hypersurfaces in Minkowski space
$t-f(x^i) = 0$ for which the perturbed data satisfies the stability
requirement (\ref{eq:ineq}). We now show that there are no such useful Cauchy hypersurfaces.
By {\em useful}, we require:
\begin{itemize}
\item That $\Sigma$ be spherically symmetric, i.e. $f(x^i)=F(r)$,
where $r^2 = x^2+y^2+z^2$. Although this requirement
does not immediately rule out more general foliations
which might satisfy (\ref{eq:ineq}), no essential generality is lost by requiring
spherically symmetry. For the purpose of exploring techniques for numerical implementation
of the algebraic-hyperbolic approach, there would be little
utility in an asymmetric
Cauchy hypersurface for simulating asymptotically flat perturbations of Minkowski space.

\item That $\Sigma$ be smooth. Smoothness at the origin requires a local
Taylor series expansion $F(r) = F(0) +\frac{r^2}{2} F'' (0) + \dots$,
where we use the notation $F'=\partial_r F$. 
For example, the spacelike hypersurface $t=C r$, $0< C <1$, has a conical
singularity at the origin whereas the hyperboloidal hypersurface
$t =F(r) =\sqrt{C^2+r^2}$ is smooth.

\item That $\Sigma$ have a complete future domain of
dependence, i.e. that it must extend from the origin $r=0$ to either spacelike
infinity or future null infinity. However, as seen below, our main result does not depend upon
this assumption.

\end{itemize}

The metric of such a hypersurface $t=F(r)$
in Minkowski space
has the $2+1$ decomposition (\ref{eq:sphdec}), with line element in spherical coordinates
given by
\begin{equation}
   d \ell^2 = A^{-2}dr^2 +r^2 q_{AB} dx^A dx^B\, , 
     \quad A =\frac{1}{\sqrt{1-F'^2}}, 
   \label{eq:dell}
\end{equation}
where  $x^A =(\theta,\phi)$ and  $q_{AB}$ is the unit sphere metric.
The corresponding curvature scalar is
$${}^{(3)}R =-\frac{2A^2 F'^2}{r^2} -\frac{4 A^4 F' F''}{r}.
$$

The future-directed unit normal to $\Sigma$ is $n_b = -A \nabla_b \big (t-F(r)\big )$
and the extrinsic curvature $k_{ab}=h_a^c \nabla_c n_b$, where $\nabla_c$ is
the covariant derivative associated with $\eta_{ab}$, is given by
$$k_{ab}=A^3F''R_a R_b +AF'rq_{ab} .
$$
Here $R_a =A (\nabla_\alpha r - F' \nabla_\alpha t)$ is the unit normal to
the $r=const$ foliation of $\Sigma$.
The background components
corresponding to (\ref{eq:TTK}) -- (\ref{eq:Ka}) are 
$K^{(TT)}_{ab} =K_a =0$
and
\begin{equation}
     \kappa = A^3 F'' \, , \quad K= \frac {2AF'}{r}.
\end{equation}

Now consider the product
\begin{equation}
        \kappa K = \frac {2A^4 F' F"}{r}. 
\end{equation}
The stability inequality (\ref{eq:ineq}) requires
\begin{equation}
                2F'F''= (F'^2)' <0.
\end{equation}
But smoothness at the origin requires $F'(0)=0$ and since  $F'^2\ge 0$
the inequality cannot be satisfied. Hence there is no suitable Cauchy hypersurface
in Minkowski space on which to base a well-posed algebraic-hyperbolic
linearized constraint problem.

\section{Discussion}

We have shown that there is no Cauchy hypersurface in Minkowski space which is useful
for numerical investigation of the algebraic-hyperbolic formulation
of the constraint problem for non-singular asymptotically flat data. This is at first surprising since such
hypersurfaces exist in a Schwarzschild spacetime, namely the $t=constant$
hypersurfaces in the ingoing Kerr-Schild form $g_{ab}=\eta_{ab}+(2m/r)\ell_a \ell_b$,
where $t$ is the inertial time for the Minkowski background metric $\eta_{ab}$.
In that case, the stability inequality
(\ref{eq:ineq}) holds everywhere outside the singularity at $r=0$. It appears that some singular structure
of the Cauchy  hypersurface is necessary for a well-posed algebraic-hyperbolic
constraint problem. In fact, for the pure Minkowski
background, the hypersurface
$$ t=\frac{Cr}{C+r}, \quad C>0 ,
$$
has extrinsic curvature components 
\begin{equation}
 \kappa =-\frac {2A^3 C^2}{(C+r)^3}\, , \quad K =  \frac{2A C^2}{r(C+r)^2} \, ,
     \quad A=\frac{(C+r)^2}{\sqrt{(C+r)^4 -C^4 }},
\end{equation}
which does satisfy the stability inequality for $r>0$.
But this hypersurface has a conical singularity
at $r=0$.

\ack

This research was supported by NSF grant PHY-1505965 to the University of Pittsburgh.
I thank I. R\'acz and C. Schell for useful communication. The problem was initiated by
discussions with L. Lehner while a guest at  the  Perimeter  Institute, which  is  supported 
by the  Government  of  Canada through NSERC and by the Province of Ontario through MEDT.


\section*{References}

\begin{thebibliography}{99}

\bibitem{lich} Lichnerowicz~A: {\it L'integration des Equations de la Gravitation Relativiste
et le Probleme des n Corps}, J. Math. Pures Appl. {\bf 23} 39 (1944) 

\bibitem{york0}  York~J W: {\it Role of conformal three-geometry in the dynamics
of gravitation}, Phys. Rev. Letters {\bf 28} 1082 (1972)

\bibitem{york1}  York~J W: {\it Covariant decompositions of symmetric tensors in the
theory of gravitation}, Ann. Inst. Henri Poincar\'e A {\bf 21} 319 (1974)

\bibitem{cook} Cook~G B: {\it Initial data for numerical relativity},
Living Rev. Relativity {\bf 3}  5 (2000)

\bibitem{gourg} Gourgoulhon~E: {\it Construction of initial data for 3+1 numerical relativity},
J. Phys.: Conf. Ser. {\bf 91} 012001 (2007)

\bibitem{racz1} R\'acz~I {\it Constraints as evolutionary systems},
Class. Quant. Grav.  {\bf 33} 015014  (2016)

\bibitem{raczwin1} R\'acz~I and Winicour~J:  {\it Black hole initial data without elliptic equations}, 
Phys. Rev. D {\bf 91}, 124013 (2015)

\bibitem{raczwin2}  R\'acz~I and Winicour~J  {\it On solving the constraints by integrating
a strongly hyperbolic system},  arXiv:gr-qc/1601.05386

\bibitem{choquet} Choquet-Bruhat~Y: {\sl General relativity and Einstein's equations},
Oxford University Press Inc., New York (2009)

\bibitem{wald} Wald~R M: \textsl{\ General relativity},
University of Chicago Press, Chicago (1984)

\bibitem{racz2} R\'acz~I: {\it Constraints as evolutionary systems},
Class. Quant. Grav. {\bf 33} 015014 (2016)

\bibitem{racz3}  R\'acz~I: {\it A simple method of constructing binary black hole initial data}, 
arXiv:gr-qc/1605.01669

\bibitem{schell} Schell~C: private communication


\end{thebibliography}
\end{document}